\documentclass[twocolumn,aps,prd,amsmath,superscriptaddress]{revtex4-1}
\usepackage{setspace}
\usepackage{color}
\usepackage{fancyhdr}
\usepackage{graphicx}
\usepackage[ansinew]{inputenc}
\usepackage{amssymb}
\usepackage{amsmath}
\usepackage{cancel}
\usepackage{float}
\usepackage{graphicx}
\usepackage{float}
\usepackage{todonotes}
\usepackage{booktabs}
\usepackage{subfigure}
\usepackage{pgfplots}
\usepackage{silence}
\usepackage{xparse}

\begin{document}

\title{Twin Stars within the SU(3) Chiral Quark-Meson Model}

\author{Andreas Zacchi}
\email{zacchi@astro.uni-frankfurt.de}
\affiliation{Institut f\"ur Theoretische Physik, Goethe Universit\"at Frankfurt, 
Max von Laue Strasse 1, D-60438 Frankfurt, Germany}

\author{Laura Tolos}
\email{tolos@ice.csic.es}
\affiliation{Institut de Ciencies de l' Espai (IEEC/CSIC), Campus Universitat Autonoma de Barcelona,
Carrer de Can Magrans, s/n, E-08193 Bellaterra, Spain}
\affiliation{Frankfurt Institute for Advanced Studies, Goethe Universit\"at Frankfurt, Ruth-Moufang-Str. 1, 60438, Frankfurt am Main, Germany}

\author{J\"urgen Schaffner-Bielich}
\email{schaffner@astro.uni-frankfurt.de}
\affiliation{Institut f\"ur Theoretische Physik, Goethe Universit\"at Frankfurt, 
Max von Laue Strasse 1, D-60438 Frankfurt, Germany}
\date{\today}

   \begin{abstract}
We present new stable solutions of the Tolman Oppenheimer Volkoff equations for quark stars using a quark matter equation of state based on the SU(3) Quark-Meson model that exhibits the onset of the chiral phase transition. These new solutions appear as two stable branches in the mass-radius relation allowing for so called twin stars, i.e. two stable quark star solutions with the same mass, but distinctly different radii. We find solutions which are compatible with causality, the stability conditions of dense matter, the astrophysical constraints of the rotation of the millisecond pulsar PSR J1748-2446ad and the 2$M_{\odot}$ pulsar mass constraint. 
   \end{abstract}

   
\maketitle        
 

   \section{Introduction}
                  
Compact stars serve as a superb laboratory to investigate the high dense equation of state (EoS), which describes the microscopic properties of the dense matter in their interior.
The recently discovered pulsars PSR J1614-2230 \cite{Demorest:2010bx,Fonseca:2016tux} and  PSR J0348+0432 \cite{Antoniadis:2013pzd} of 2$M_{\odot}$ have revived the discussion on the dense phases inside neutron stars. In fact, depending on the type of matter in the interior of these compact objects, several possibilities for the nature of neutron stars have been postulated: strange quark stars, normal neutron stars or hybrid stars. Compact stars entirely made of deconfined quark matter (besides maybe a small layer of a crust of nuclei) are called
strange quark stars \cite{Bodmer:1971we,Witten:1984rs,Weber:2004kj}, whereas compact objects made of hadrons are referred to normal neutrons stars \cite{Lattimer:2006xb}. Hybrid stars are compact stars with a core consisting of quark matter and outer layers of hadronic matter. For a recent review see \cite{Buballa:2014jta}. 

Two different equations of state are therefore needed to describe hybrid stars, that is, one for the quark matter interior and another one for the hadronic outer layers. Depending on the features of the transition between the inner and outer parts of hybrid stars
\cite{Glendenning:1998ag,Schertler:1998cs,SchaffnerBielich:2002ki,Alford:2013aca,Alford:2015dpa}, a twin star configuration might arise. Twin stars appear as the mass-radius relation exhibits two stable branches with similar masses, that is, a third family of compact stars appears with alike masses as the second family branch of normal neutron or strange quark stars.

Stable twin star solutions have been discussed in the literature over the past years  \cite{Glendenning:1998ag,Schertler:1998cs,SchaffnerBielich:2002ki,Alford:2013aca,Alford:2015dpa,Blaschke:2015uva}. In particular, 
twin stars respecting the 2$M_{\odot}$ constraint have been recently studied based on a density-dependent hadronic EoS for the star outer crust and a NJL-model 
type EoS for the quark matter phase within the star core \cite{Benic:2014jia}. 

In this work we obtain an EoS for quark matter that exhibits a genuine transition from a chirally non-restored phase to a restored one, allowing for the existence of twin star solutions with only quark matter. The EoS results from the SU(3) Chiral Quark-Meson Model, which is based on the linear $\sigma$-model \cite{Lenaghan:2000ey,Schaefer:2008hk} and couples mesons and quarks by respecting chiral symmetry. Whereas this model has been previously used within the context of compact stars  \cite{Zacchi:2015lwa,Zacchi:2015oma}, in the present work we go beyond these previous calculations. By scanning the allowed parameter range of the SU(3) Chiral Quark-Meson model - that is the mass of the sigma meson $m_{\sigma}$, the  vector coupling $g_{\omega}$ and the vacuum pressure $B^{1/4}$ - we find two stable branches for certain values of the vector coupling and the sigma mass, while keeping the vacuum pressure below $B \lesssim 100$ MeV. Some of these twin star configurations are compatible with recent 2$M_\odot$ observations and the mass-radius constraints coming from the rotation of the millisecond pulsar PSR J1748-2446ad \cite{Hessels:2006ze}, while still satisfying causality and the stability conditions of dense matter.

This paper is organized as follows. In Section \ref{formalism} we present the SU(3) Chiral Quark-Meson model and the constraints from causality and rotation on the mass-radius relation. Next, in Section \ref{results} we show our results for twin stars by varying the  parameters of the model, that is, the sigma meson $m_{\sigma}$, the vector coupling $g_{\omega}$ and the vacuum pressure $B^{1/4}$. Moreover, we examine the stability conditions of these solutions and determine those that are compatible with the 2$M_{\odot}$ observations, as well as the  rotation of the millisecond pulsar PSR J1748-2446ad. Finally, in Section \ref{conclusions} we present our conclusions.


   \section{Compact Stars within the SU(3) Chiral Quark-Meson Model}
   \label{formalism}


\subsection{The SU(3) Chiral Quark-Meson Model}  
The SU(3) Chiral Quark-Meson model is based on the linear $\sigma$-model \cite{Lenaghan:2000ey,Schaefer:2008hk} and couples mesons and
quarks by means of chiral symmetry. The Lagrangian $\mathcal{L}$ of the SU(3) Chiral Quark-Meson model reads  \cite{Scavenius:2000qd}
\begin{eqnarray}
\label{qmlagrangian} 
\mathcal{L}&=&\bar{\Psi}\left(i\cancel{\partial}-g_\varphi\varphi-g_v\gamma^\mu V_\mu\right)\Psi\\ \nonumber &+&tr\left[(\partial_{\mu}\varphi)^{\dagger}(\partial^{\mu}\varphi)\right]+
tr\left[(\partial_{\mu} V)^{\dagger}(\partial^{\mu} V)\right]\\ \nonumber
&-&\lambda_1[tr(\varphi^{\dagger}\varphi)]^2-\lambda_2
tr(\varphi^{\dagger}\varphi)^2-m_0^2(tr(\varphi^{\dagger}\varphi))\\ \nonumber
&-&m_v^2 (tr(V^{\dagger}V))
+tr[\hat{H}(\varphi+\varphi^{\dagger})]+c\left(\det(\varphi^{\dagger})+
\det(\varphi)\right) , 
\end{eqnarray}
for $SU(3)_{\rm L} \times SU(3)_{\rm R}$ chiral symmetry incorporating the scalar ($\varphi$) and
vector ($V_\mu$) meson nonet.  Here, $m_v$ stands for the vacuum mass of the 
vector mesons $\omega$, $\rho$ and $\phi$. The quantities $\lambda_1$, $\lambda_2$, $m_0$, and $c$ 
are the standard vacuum parameters of the linear $\sigma$ model 
\cite{Lenaghan:2000ey} that depend on the meson masses and the sigma meson mass, $m_{\sigma}$. The matrix 
$\hat{H}$ describes the explicit breaking of chiral symmetry. The 
quarks couple to the meson fields via Yukawa-type interaction terms with the 
coupling strengths $g_\varphi$ for scalar and $g_v$ for vector mesons, respectively.

The grand canonical potential $\Omega$ in the mean field approximation at vanishing temperature 
can be derived via the path integral formalism. The 
coupled equations of motion of the meson fields are then determined via the 
derivative of the grand potential to the respective fields  (see \cite{Zacchi:2015lwa} for details).

The pressure $p=-\Omega$ is  given by
   \begin{eqnarray} \nonumber
  p=&-&\frac{\lambda_1}{4}(\sigma_n^2+\sigma_s^2)^2-\frac{\lambda_2}{4}(\sigma_n^4
  +\sigma_s^4) 
  -\frac{m_0^2}{2}(\sigma_n^2+\sigma_s^2) \nonumber \\
  &&+  \frac{\sigma_n^2\sigma_s}{2\sqrt{2}}c +h_n\sigma_n+h_s\sigma_s-B \nonumber \\
 &&  + \frac{1}{2}\left(m_{\omega}^2\omega^2+m_{\rho}^2\rho^2+m_{\phi}^2\phi^2
  \right) \nonumber  \\  
  &&-\frac{3}{\pi^2}\sum_{f=u,d,s}\int_0^{k_F^f}dk\cdot k^2\left(\sqrt{k^2
  +\tilde{m}_{f}^2}-\tilde{\mu}_{f}\right) .
  \label{press} 
    \end{eqnarray}
Here $B^{1/4}$ is a vacuum energy term. The index $n$  stands for the nonstrange up ($u$) and down ($d$) quarks, and $s$ for strange quarks. The quantities
$h_n$ and $h_s$ are the explicity symmetry breaking terms resulting from $\hat{H}$. 
The effective quark mass  is given by $\tilde{m}_f=g_{\varphi} \sigma_{f}$ and $f$ indicates the flavour of the respective quark. The flavour dependent chemical potential is $\tilde{\mu}_f$. The integral in Eq.~(\ref{press})  for the different quark flavours runs up to the corresponding Fermi momentum, $k_F^f$.

Once the pressure is known, the energy density can be calculated from the relation, $\Omega=\epsilon +\sum_f \mu_f n_f$, where $n_f=(k_F^f)^3/ \pi^2$ is the density associated with each quark flavour. The equation of state (EoS) is then given by $p(\epsilon)$. 


   \subsubsection{Parameters of the SU(3) Quark-Meson Model}\label{parameters}

A detailed analysis of the different parameters of the SU(3) Quark-Meson model can be found in \cite{Lenaghan:2000ey,Schaefer:2008hk,Zacchi:2015lwa}.
In this section we show the four free parameters that can be varied in order to scan the allowed parameter range of the SU(3) Chiral Quark-Meson model.
   \begin{enumerate}
 \item The constituent quark mass $m_q$ determines the scalar Yukawa-like coupling constants $g_{\varphi}$ for the 
       nonstrange, $g_n$, and strange quark, $g_s$, via the Goldberger-Treiman 
       relation. That is, $g_n=\frac{m_q}{f_{\pi}}$, where $f_{\pi}$ is the vacuum expectation value of the sigma field,
       and $g_s=g_n\sqrt{2}$ from SU(3) symmetry. In this paper we fix $m_q=300$~MeV, which is roughly $1/3$ of the nucleon mass.
 \item The vector coupling is independent of the constituent quark mass. It is usually 
       varied in the scale of the scalar coupling, $g_{\omega}\sim g_n \sim 3$, (see \cite{Beisitzer:2014kea,Zacchi:2015lwa,Zacchi:2015oma}). 
       The strange coupling of the
       $\phi$-meson is  fixed by SU(3) symmetry constraints,  $g_{\phi}=\sqrt{2}g_{\omega}$. For the analysis of twin stars we will use $g_{\omega} \lesssim 10$.
 \item The mass of the sigma meson $m_{\sigma}$ is not well determined experimentally. Usually,  the sigma meson, which is the chiral partner of the pion,
       is identified with the experimentally measured resonance $\rm{f_0(500)}$, 
       which is rather broad, $400 \leq m_{f_0} \leq 600$~MeV \cite{Agashe:2014kda}. 
       Also, in Ref.~\cite{Parganlija:2012fy} it was demonstrated 
       that within an extended quark-meson model that includes vector and 
       axial-vector interactions, the resonance $\rm{f_0(1370)}$ could be 
       identified as the scalar state.  
       Here we vary $400 \leq m_{\sigma} \leq 800$~MeV for finding twin star solutions.
 \item  The value of $B^{1/4}$ determines whether strange quark matter 
       has a lower ground state than hadronic matter in the confined phase following the Bodmer-Witten hypothesis \cite{Bodmer:1971we,Witten:1984rs}. 
      We will explore also values of $B^{1/4}$ even below $\sim100$~MeV in order to study the possibility of twin stars. 
              
   \end{enumerate}


\subsection{Mass-Radius Relation and Constraints from Causality and Rotation}    
\label{constraints}

Once the EoS is known, the mass $M$ and the corresponding radius $R$ of the compact star are obtained from solving the Tolman-Oppenheimer-Volkoff (TOV) equations
  \begin{eqnarray} \label{eq:tov_1}
&&\frac{dM(r)}{dr}=4\pi \epsilon (r) r^2  , \nonumber \\
&&\frac{dP(r)}{dr}=- \frac{\left[p(r)+\epsilon(r)\right] \left[M(r)+4\pi r^3 p(r) \right]}{r(r-2M(r))} , 
\label{tov}
   \end{eqnarray}
 in units  $c=G=1$, where $r$ is the radial coordinate. To solve these equations one needs to specify the enclosed mass and the pressure at the center of the star, $M(r=0)=0$ and $p(r=0)=p_c$, while the energy density is taken from the assumed EoS. The integration of the TOV equations over the radial coordinate ends when $p(r=R)$=0. 

Several constraints have to be fullfilled by the mass-radius relation of a compact object. In Ref. \cite{Lattimer:2006xb} a rigorous causal limit for normal neutron stars has been presented, being $R \geq 2.87 M$. 
Also, neutron stars are usually observed as pulsars, rotation powered neutron stars. The rotation frequency is limited by the Keplerian (or mass-shedding) frequency, which is a rotational limit obtained when the equatorial surface velocity equals the orbital speed just above the surface. Lattimer and Prakash determined an empirical formula for the  mass-shedding limit for an arbitrary neutron star mass, as long as the mass is not close to the maximum mass \cite{LP04,Lattimer:2006xb} 
\begin{equation}
P_{\rm min}\simeq (0.96\pm0.03)\left(\frac{M_{\odot}}{M_{nr}}\right)^{1/2}\left(\frac{R_{nr}}{10 \rm{km}}\right)^{3/2} \ \ {\rm ms} ,
\label{kep_freq}
\end{equation}
with $M_{nr}$ and $R_{nr}$ referring to the non-rotating mass and radius from the TOV equations. Eq.~(\ref{kep_freq}) can be used as an estimate to limit masses and radii for stars made of quarks \cite{Weissenborn:2011qu}. In order to constrain the allowed mass-radius for compact stars, we will follow previous works \cite{Lattimer:2006xb,LP04} and impose that the maximum spin rate allowed is given by the most rapidly rotating compact star observed up to now, the pulsar PSR J1748-2446ad, with a spin rate of 716 Hz \cite{Hessels:2006ze}.

 
   \section{Signatures of Twin stars}
   \label{results}
 
 
  \subsection{Influence of the vacuum pressure}

 \begin{figure}[t]
\center
\includegraphics[width=\columnwidth, height=\columnwidth]{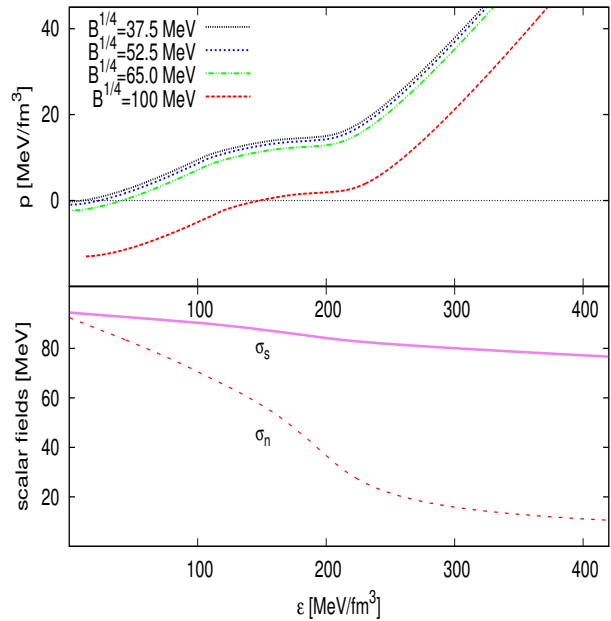}
\caption{\textit{The EoS (upper plot) as well as the non-strange and strange scalar fields (lower plot) as a function of the energy density for the parameter set $m_q=300$~MeV, $m_{\sigma}=600$~MeV, $g_{\omega}=4$ and different values of the vacuum pressure $B^{1/4}$. }}
\label{gw4_1}
\end{figure}
 
 In order to study the possible existence of twin stars, we start by varying the value of the vacuum pressure $B^{1/4}$. The change of the value of the vacuum pressure modifies the stiffness of the EoS, and, hence, the maximum mass \cite{Witten:1984rs}, while the meson fields are independent of $B^{1/4}$ \cite{Schertler:1998cs,Glendenning:1998ag,Zacchi:2015lwa,Zacchi:2015oma}.\\
 
In Fig.~\ref{gw4_1} we show the EoS (upper plot) as well as the non-strange and strange scalar fields (lower plot) as a function of the energy density for the parameter choice $m_q=300$~MeV, $m_{\sigma}=600$~MeV, $g_{\omega}=4$ and  different values of the vacuum pressure $B^{1/4}$. As expected, we find that small values of $B^{1/4}$ $\sim$ 30-70 MeV stiffen the EoS, while the meson fields are unaffected by this change. The non vanishing value of the energy density at zero pressure, typical for the EoS of self-bound stars such as strange quark stars, gets closer to zero as we reduce the value of $B^{1/4}$. As this happens, a smooth non-linear behaviour of the pressure with the energy density for $100\rm{MeV/fm^3} \leq \epsilon \leq 210\rm{MeV/fm^3}$ becomes noticeable at positive pressures.  This non-linearity results from a crossover-type chiral phase transition taking place in this energy range, as seen in the behaviour of the scalar fields in the lower plot of Fig.~\ref{gw4_1}. Moreover, this non-linearity for $B^{1/4} \lesssim 100$ MeV separates two different energy-density regions with positive, increasing pressure, that gives rise to the appearance of two separate stable branches in the mass-radius relation and, hence,  the existence of twin stars, as we shall see next.

\begin{figure}[t]
\center
\includegraphics[width=\columnwidth]{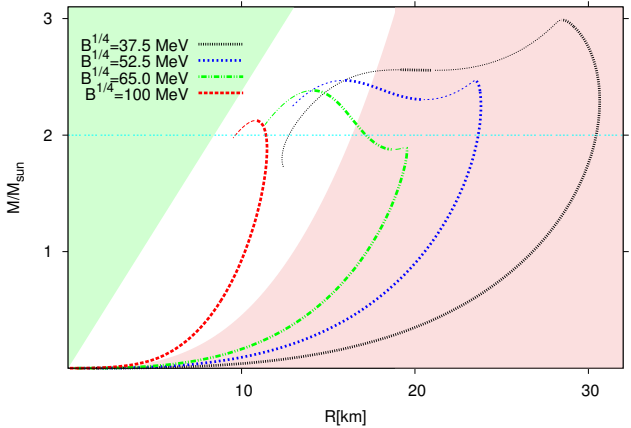}
\caption{\textit{The mass-radius relation for the parameter set $m_q=300$~MeV, $m_{\sigma}=600$~MeV, $g_{\omega}=4$ and different values of the vacuum pressure $B^{1/4}$. 
The upper left shaded region is excluded due to causality whereas the lower shaded area on the right-hand side is forbidden by the rotation of the millisecond pulsar PSR J1748-2446ad. The horizontal line indicates the 2$M_{\odot}$ limit. Thick lines in the mass-radius relation represent stable configurations, whereas thin ones correspond to unstable solutions.  }}
\label{mrr_1}
\end{figure}

The mass-radius relation for $m_q=300$~MeV, $m_{\sigma}=600$~MeV, $g_{\omega}=4$ and different values of  $B^{1/4}$  is depicted in Fig.~\ref{mrr_1}.  The upper left shaded region of Fig.~\ref{mrr_1} is excluded due to causality whereas the lower shaded area on the right-hand side is ruled out by the rotation of the millisecond pulsar PSR J1748-2446ad, as described in Sec.~\ref{constraints}, using Eq.~(\ref{kep_freq}). The horizontal line indicates the 2$M_{\odot}$ limit. 

We first note that heavier stars with larger radii are obtained for low values of $B^{1/4}$ as compared to the $B^{1/4}=100$ MeV case, due to the fact that the stiffness of the EoS increases. The stiffer the EoS is, the larger the masses are and also the larger the radii become. Moreover, depending on the value of the vacuum pressure, we find different scenarios for the mass-radius relation. The smallest value $B^{1/4}=37.5$~MeV gives rise to two stable mass-radius branches, the twin star configuration. For that particular value of $B^{1/4}$, there exist two maximum masses, $M^{max}_1$ and $M^{max}_2$, being $M^{max}_1 \gtrsim M^{max}_2$.  For the case of  $B^{1/4}=52.5$~MeV, we obtain $M^{max}_1 \simeq M^{max}_2$, while for $B^{1/4}=65$~MeV we find that $M^{max}_1 \lesssim M^{max}_2$. The value $B^{1/4}=100$~MeV does not yield to a second stable branch, thus, only one maximum mass is obtained. The values of the central pressure, central energy density, maximum mass and radius for these maxima are shown in Table \ref{terence_hill}. 

The appearance of two separate stable branches in the mass-radius relation comes from the chiral phase transition that causes the non-linear behaviour of the EoS for values of $100\rm{MeV/fm^3} \leq \epsilon \leq 210\rm{MeV/fm^3}$ (see Fig.~\ref{gw4_1}). The physical interpretation is the following: a first branch of quark matter in a chirally broken phase develops until the chiral phase transition sets in. After an unstable region, a new stable second branch of chirally restored quark matter emerges. 

We also find that, for the given values of $m_{\sigma}$ and $g_{\omega}$, the chiral phase transition is a crossover and not first order. This is in contrast to the studies on twin stars coming from hybrid configurations, where a first-order phase transition takes place \cite{Glendenning:1992vb}, but supports the robustness of the twin star solution with regard to smoothing the phase transition found in \cite{Benic:2014jia}, see also \cite{Alvarez-Castillo:2014dva}.
As we increase the value of $B^{1/4}$, the phase transition is moving from the outer parts to the inner core of the quark star. For $B^{1/4} \gtrsim100$~MeV, the chiral phase transition is shifted to negative pressure (Fig.~\ref{gw4_1}) and, thus, no stable Twin Star solutions are possible, as seen in Fig.~\ref{mrr_1}. In this case, we recover the $M \propto R^3$ relation for self-bound stars. Note that only the mass-radius branches inside the non-shaded areas are allowed by causality and the rotation of PSR J1748-2446ad, while the 2$M_{\odot}$ limit sets a lower limit for the maximum masses.

\begin{table*}
\begin{center}
\begin{tabular}{|c|c|c||c|c||c|c||c|c||c|c|}
\hline\hline 
\multicolumn{3}{|c||}{Parameters}  & \multicolumn{2}{r}{First}  & \multicolumn{2}{l||}{maximum}  & \multicolumn{2}{r}{Second} & \multicolumn{2}{l|}{maximum}   \\
\hline
\hspace{.2cm} $B^{1/4}$ \hspace{.2cm}& \hspace{.2cm} $m_{\sigma}$ \hspace{.2cm} & $g_{\omega}$ & \hspace{.2cm} $p_c$ \hspace{.1cm} & $\epsilon_c$ &  $M/M_{\odot}$ & \hspace{.2cm} $R$(km)  \hspace{.2cm} & \hspace{.2cm} $p_c$ \hspace{.2cm} &\hspace{.1cm} $\epsilon_c$ \hspace{.2cm} & \hspace{.2cm} $M/M_{\odot}$ \hspace{.2cm} & $R$(km) \\
\cline{1-11}
37.5 & 600 & 4 & 13.6 & 151.2 & 2.99 & 28.46 & 141.9 & 577.6 & 2.56 & 19.18  \\
52.5 & 600 & 4 & 12.99 & 153.75 &  2.46 & 23.54 & 209.65 & 736.45 & 2.46 & 15.94  \\ 
65 & 600 & 4 &13.55& 210.87 &  1.89 & 19.44 &271 &873.91&2.38 & 14.07 \\ 
100 & 600 & 4 & 404.04 & 1176.57 & 2.12 & 10.83 & - & - & - & -
\\ \hline\hline
\end{tabular}
\caption{\textit{The central pressure $p_c$ and the corresponding central energy density $\epsilon_c$ as well as the masses and radii for the first and second maximum for $m_q=300$~MeV, $m_{\sigma}=600$~MeV, $g_{\omega}=4$ and for different values of the vacuum pressure $B^{1/4}$.}}
\label{terence_hill}
\end{center}
\end{table*}

\begin{figure}[t]
\center
\includegraphics[width=\columnwidth]{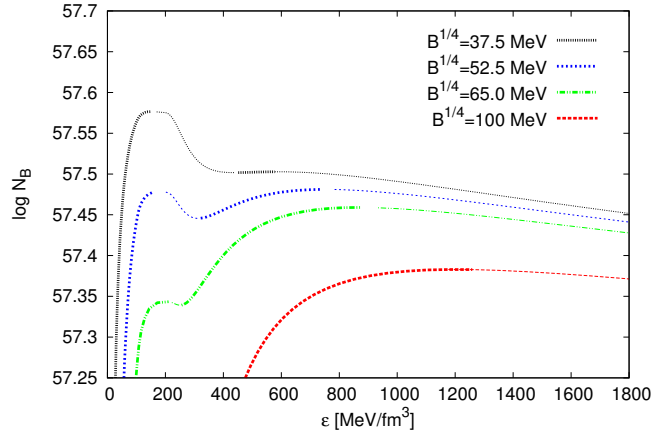}
\caption{\textit{The baryon number $N_B$ as a function of the energy density for the parameter set $m_q=300$~MeV, $m_{\sigma}=600$~MeV, $g_{\omega}=4$ and various values of the vacuum pressure $B^{1/4}$. Thick lines represent stable configurations, whereas thin lines correspond to unstable solutions. }}
\label{nb}
\end{figure}

At this stage it is pertinent to ask whether the collapse of a star in the first branch (or second family) into a star with the same mass, belonging to the second branch (or third family) is possible by introducing small perturbations that conserve the baryon number, such as compression or temperature fluctuacions. In order to assess this possibility it is necessary to show that, on the one hand, a star in the first branch has the same baryon number as one in the second branch and moreover, on the other hand, the binding energy per baryon is higher for the twin in the second branch, in order to have a more stable configuration.

The baryon number $N_B$ can be calculated as in \cite{Glendenning:1998ag}
\begin{eqnarray} \label{eq:tov_3}
 N_B&=&\frac{4\pi}{3}\int_0^R \left(1-\frac{2M(r)}{r}\right)^{-1/2} \rho_q(r) r^2 dr  , 
\end{eqnarray}
where $\rho_q(r)$ is the quark number density within the star. Then, the baryonic mass is given by  $M_B= 3 N_B m_q $, with $m_q$ being the constituent quark mass, and the binding energy per baryon is defined as  $E_B=(M_B-M)/N_B$.

In Fig.~\ref{nb} we show the logarithm of the baryon number as a function of the energy density for  $m_q=300$~MeV, $m_{\sigma}=600$~MeV, $g_{\omega}=4$ and the previously chosen values of the vacuum pressure $B^{1/4}$. We observe that for $B^{1/4} \lesssim 100$~MeV some stars of the two stable branches have the same mass (twin partners) with the same baryon number. However, our calculations indicate that the condition for collapse coming from the binding energy is fulfilled only by the  twin partners for $B^{1/4}=52.5$~MeV. Moreover, only for $B^{1/4}=52.5$~MeV it is possible to have few stable twin partners in the second branch that sit in the region that is not excluded by causality and rotation, as seen in Fig.~\ref{mrr_1}. 

\begin{figure}[t]
\center
\includegraphics[width=\columnwidth]{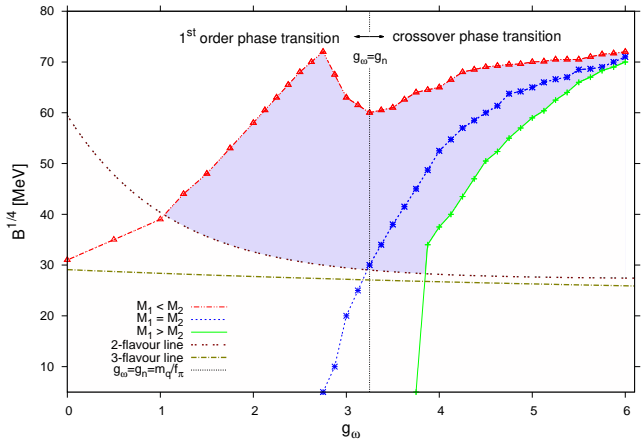}
\caption{\textit{Contour lines in the $g_{\omega}$ versus $B^{1/4}$ plane indicating Twin Star solutions. Two stable branches with maxima $M^{max}_1<M^{max}_2$ appear when crossing the upper line from above. The middle line indicates stars with $M^{max}_1 \simeq M^{max}_2$. The lower line shows the limit where $M^{max}_1>M^{max}_2$. Above the upper line and below the lower line no twin stars are possible. The vertical line $g_{\omega}=g_n$ indicates the approximate boundary between first order- and crossover phase transitions. The two- and three flavour lines provide the stability conditions for dense matter. Twin stars within the shaded area fullfill the stability constraints. }}
\label{walker_tex_ass_ranger}
\end{figure}
\begin{figure}[t]
\center
\includegraphics[width=\columnwidth]{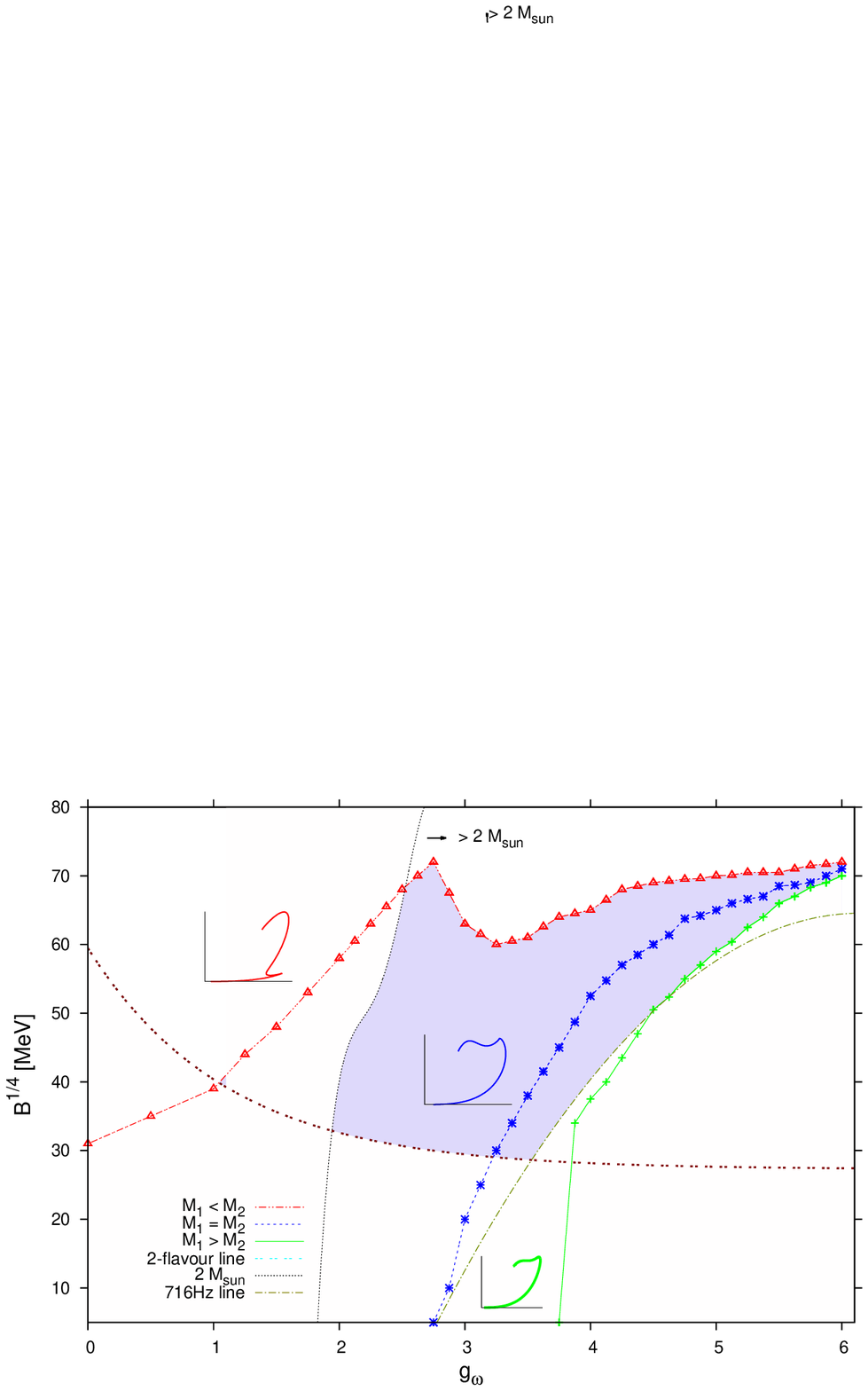}
\caption{\textit{Same contour lines in the $g_{\omega}$ versus $B^{1/4}$ plane indicating Twin Star solutions, as in Fig.~\ref{walker_tex_ass_ranger}, but  including the two-flavour line, the 2 $M_{\odot}$ limit and also showing the region excluded by the rotation of the millisecond pulsar PSR J1748-2446ad.  Twin stars within the shaded area fullfill all these constraints. }}
\label{other}
\end{figure}

\subsection{Influence of  the Vacuum Pressure and the Vector Coupling Constant}

In order to further investigate the conditions for the existence twin stars, we consider not only the variation of the vacuum pressure, $B^{1/4}$, but also the vector coupling constant, $g_{\omega}$, within reasonable margins. For the sigma mass we keep the previous standard value of $m_{\sigma}=600$ MeV. 

Fig.~\ref{walker_tex_ass_ranger} and Fig.~\ref{other} depict the $g_{\omega}-B^{1/4}$ parameter plane at $m_{\sigma}=600$~MeV. By crossing the upper line from above we obtain two stable branches with two maxima being related as $M^{max}_1 \lesssim M^{max}_2$. At the middle line the two stable branches have $M^{max}_1\simeq M^{max}_2$, whereas the second stable branch disappears for small values of $B^{1/4}$ and high values of $g_{\omega}$ below the lower line. 

From Fig.~\ref{walker_tex_ass_ranger} we see that, for a fixed value of $B^{1/4}$, at high values of $g_{\omega}$ the twin star solutions disappear as there is no chiral phase transition. Low $g_{\omega}$ requires $30 \lesssim B^{1/4} \lesssim 70$~MeV for twin stars to appear. The vertical line indicates the approximate boundary between first order- and crossover phase transitions, that is given by $g_{\omega} \sim g_n$ for $m_{\sigma}=600$ MeV.

In Fig.~\ref{walker_tex_ass_ranger} we also show the two-flavor line resulting from the stability condition of nuclear matter and the three-flavor line coming from the Bodmer-Witten hypothesis of stable strange quark matter \cite{Bodmer:1971we,Witten:1984rs}. The region above the two-flavor line indicates that two-flavor quark matter is not more stable than ordinary nuclear matter, meaning that the most stable known element in nature, $^{56}{\rm Fe}$, cannot decay into two-flavor quark matter. The area below the three-flavor line concerns the Bodmer-Witten hypothesis, i.e. strange quark matter is more stable than ordinary nuclei. Since there is no overlap between both regions, absolutely stable quark matter is not realizable. 

There are further constraints to the allowed parameter range for twin stars  coming from astrophysical observations. In Fig.~\ref{other} we show that, for $g_{\omega} \gtrsim 2$, the masses of the stars become larger than the 2$M_{\odot}$ limit. The parameter range for Twin Star configurations lie inside the area allowed by the rotation of PSR J1748-2446ad. These twin stars have radii $12$ Km $\lesssim R \lesssim 22 $ Km, which are larger than the recent determinations below of $R\sim 11$~km (see the discussion in Ref.~\citep{Ozel:2016oaf}).


\subsection{Influence of the Sigma Mass}

In this section we comment on how changes in the sigma mass affect the existence of twin stars. Note that by increasing the $m_{\sigma}$, the chiral phase transition  
becomes a smooth crossover and the corresponding EoS gets softer.  

As we reduce the sigma mass to $m_{\sigma}=400$ MeV, bigger values of  $g_{\omega}$ are needed ($6 \lesssim g_{\omega} \lesssim 10$)  in order to find twin star solutions as the ones depicted in Fig.~\ref{walker_tex_ass_ranger} for the $g_{\omega}-B^{1/4}$ parameter range. This is due to the fact that a reduction of the sigma mass implies an increase of attraction which is compensated by an increased repulsion by augmenting the value of $g_{\omega}$. For lower sigma mass and higher value of $g_{\omega}$, the EoS becomes stiffer and, thus, the masses and radii are larger than for the $m_{\sigma}=600$ MeV case. Whereas the 2$M_{\odot}$ constraint is satisfied, these Twin Star configurations are ruled out by the rotational constraint imposed by PSR J1748-2446ad. 

On the contrary, for higher sigma masses of $m_{\sigma}=800$ MeV, twin stars appear for $g_{\omega} \lesssim 2$, thus leading to Twin Star configurations with smaller masses and radii than for the case of $m_{\sigma}=600$ MeV. Although the rotational constraint set by PSR J1748-2446ad  is fulfilled, the limit of 2$M_{\odot}$ can be hardly reconciled by our Twin Star solutions for $m_{\sigma}=800$ MeV. Only twin stars of 2$M_{\odot}$ are possible for $1 \lesssim g_{\omega} \lesssim 2$ and $B^{1/4} \lesssim 20$ MeV. 


\section{Conclusions}
\label{conclusions}

We study the possibility of twin stars within the SU(3) Chiral Quark-Meson model.  We find that the appearance of two stable branches in the mass-radius relation and, hence, the existence of twin stars is related to the onset of the chiral phase transition in quark matter.  

For the analysis we vary the SU(3) parameters of the model, that is, the vacuum pressure, the vector coupling and the sigma mass. For a vacuum pressure below $B \sim 100$ MeV the EoS for quark matter exhibits a genuine transition from a chirally broken phase to  a restored one and, thus, allows for the existence of twin stars. The interplay between the vector coupling and the sigma mass is then crucial for having Twin Star solutions that fulfill causality and the stability conditions of dense matter \cite{Bodmer:1971we,Witten:1984rs} as well as the astrophysical constraints coming from the rotation of the millisecond pulsar PSR J1748-2446ad  \cite{Hessels:2006ze} and the 2$M_{\odot}$ constraint \cite{Demorest:2010bx,Antoniadis:2013pzd,Fonseca:2016tux}. 

For $m_{\sigma}=600$ MeV, twin stars fulfill the 2$M_{\odot}$ limit for $g_{\omega} \gtrsim 2$ whereas values of $30$ MeV $\lesssim B^{1/4} \lesssim 70$ MeV are needed to satisfy the stability conditions of dense matter. The  constraint from rotation of PSR J1748-2446ad further reduces the allowed parameter region. Smaller values of $m_{\sigma}$ are ruled out by the rotational constraint while bigger values of $m_{\sigma}$ imply masses below the 2$M_{\odot}$ observations. 

The radii of the twin star configurations turn out to be above 12 Km. Recent determinations of stellar radii suggest values of 11 km \citep{Ozel:2016oaf}. 
With space missions such as  NICER (Neutron star Interior Composition ExploreR) \citep{2014SPIE.9144E..20A}, 
high-precision X-ray astronomy will be able to offer precise measurements of masses and radii, with a 1 Km resolution \citep{Watts:2016uzu}. The discovery of two stars with the same masses but different radii could be indeed a signal of the existence of twin stars implying the presence of a phase transition in dense matter.

\begin{acknowledgments}
The authors thank Matthias Hanauske 
for discussions throughout the whole project. 
AZ is supported by the Helmholtz Graduate School
for Heavy-Ion Research (HGS-HIRe), the Helmholtz Research 
School for Quark Matter (H-QM) and the Stiftung Giersch.
LT is supported by from the Ram\'on y Cajal research programme,
FPA2013-43425-P Grant from Ministerio de Economia y Competitividad (MINECO) and NewCompstar COST Action MP1304.
\end{acknowledgments}

\bibliography{biblio}
\bibliographystyle{apsrev4-1}
\end{document}